\documentstyle[fleqn]{tp}

\input psfig

\def\gs{\mathrel{\raise1.16pt\hbox{$>$}\kern-7.0pt
\lower3.06pt\hbox{{$\scriptstyle \sim$}}}}
\def\ls{\mathrel{\raise1.16pt\hbox{$<$}\kern-7.0pt
\lower3.06pt\hbox{{$\scriptstyle \sim$}}}}
\newcommand{\himpc}{{\hbox {$h^{-1}$}{\rm Mpc}} }
\def\himsun{{h^{-1}M_\odot}}

\def\pppm{\rm P^3M}

\begin{document}

\twocolumn[
\title{Density profiles and clustering of dark matter halos}
\author{Y.P. Jing \\
{\it Research Center for the Early Universe }\\
{\it    University of Tokyo, Bunkyo-ku, Tokyo 113-0033, Japan}}
\endabstract]

\markboth{Y.P. Jing}{Density profiles and clustering of dark matter halos}

\section {N-body simulations with 17 million particles}

Dark Matter (DM) halos carry important information about the formation
of galaxies and of many other extragalactic objects.  In particular,
the density profile and the clustering of DM halos are two
very important ingredients for understanding many observations of
galaxies. In this talk, I present our new results for these two
quantities from a large set of high quality N-body simulations.

The simulations were generated with a vectorized $\pppm$
(i.e. Particle Particle Particle Mesh) code on the the supercomputer
VPP300/16R at the National Astronomical Observatory of Japan. Each
simulation uses $256^3$ simulation particles. The code adopts the
standard $\pppm$ algorithm. Twenty one simulations are generated for three
representative cold dark matter (CDM) models of galaxy formation, and
eighteen for six scale-free cosmologies. The CDM models are specified
completely as regards the DM distribution by the density parameter
$\Omega_0$, the cosmological constant $\lambda_0$, the shape $\Gamma$
and the normalization $\sigma_8$ of the linear power spectrum. The
scale-free models assume an Einstein--de Sitter universe (i.e.
$\Omega_0=1$ and $\lambda_0=0$) and a power-law $P(k)\propto k^n$ for
the linear density power spectrum. Table~1 summarizes the physical and
simulation parameters used for these simulations.

{\small
\begin{table*}
\caption[]{\hspace{4pt} List of simulations\\
1. All simulations use $256^3$ particles, except the low density
models with power-law $P(k)$ LSF1 to LSF4 use
$200^3$ particles;\\
2. Particle mass $m_p$ in CDM models is units of $\himsun$;\\
3. Box size $L$ in CDM models is units of $\himpc$;\\
4. The rightmost column is the number of realizations (samples);
}
\vspace{6pt}
\begin{center}
\begin{tabular}{cccccccccc}
\hline\hline\\[-6pt]
Model & $\Omega_0$ &  $\lambda_0$  
&  $\Gamma$ &   $\sigma_8$ & $m_p$& $L$ & Num.\\ 
[4pt]\hline \\[-6pt]
CDM1 & 1.0  & 0.0 & 0.5 & 0.6 & $1.7\times 10^{10}$ & 100 & 3\\
CDM2 & 0.3  & 0.0 & 0.25 & 1.0 &$5.0\times 10^{9}$& 100& 3\\
CDM3 & 0.3  & 0.7 & 0.20 & 1.0 &$5.0\times 10^{9}$&100&3\\
[4pt]\hline \\[-6pt]
CDM4 & 1.0  & 0.0 & 0.5 & 0.6 & $4.6\times 10^{11}$ & 300 & 4\\
CDM5 & 0.3  & 0.0 & 0.25 & 1.0 &$1.3\times 10^{11}$& 300& 4\\
CDM6 & 0.3  & 0.7 & 0.20 & 1.0 &$1.3\times 10^{11}$&300&4\\
[4pt]\hline \\[-10pt]

Model & $\Omega_0$ &  $\lambda_0$  
& $n$&&&& Num.\\ 
[4pt]\hline \\[-6pt]
SF1& 1.0  & 0.0 &1.0&&&&3\\
SF2& 1.0  & 0.0 &0.0&&&&3\\
SF3& 1.0  & 0.0 &$-0.5$&&&&3\\
SF4& 1.0  & 0.0 &$-1.0$&&&&3\\
SF5& 1.0  & 0.0 &$-1.5$&&&&3\\
SF6& 1.0  & 0.0 &$-2.0$&&&&3\\
LSF1& 0.1  & 0.9 &$-1.0$&&&&2\\
LSF2& 0.1  & 0.0 &$-1.0$&&&&2\\
LSF3& 0.1  & 0.9 &$-2.0$&&&&2\\
LSF4& 0.1  & 0.0 &$-2.0$&&&&2\\
\hline
\end{tabular}
\label{tab:1}
\end{center}
\end{table*}
}

This set of simulations has covered a large parameter space, which is
often important in many studies. Each of our individual
simulations has similar mass and force resolutions to that the Virgo
Consortium (Gus August in these proceedings) has recently obtained
(except for their Hubble-volume simulation which has many more
particles but still poorer absolute mass resolution than the CDM
simulations presented here), but our sample is much larger. The CDM
simulations have been used to study the implications of the strong
clustering discovered by Steidel et al. (1998) of the Lyman Break
galaxies at high redshift (Jing \& Suto 1998).  Many works based on
these high quality simulations are being done. As an incomplete
list, we are 1) studying the DM distribution in the strongly
clustering regime; 2) testing the Press-Schechter extension theories;
3) studying redshift distortion effect and investigating its potential
to measure the cosmological parameters; 4) analyzing internal
structures of DM halos; 5) investigating galaxy formation by
hand-inputing gas physics; 6) applying the simulations to real
observations. 

\section{The density profile of the dark matter halos}
\begin{figure*}
\centering\mbox{\psfig{figure=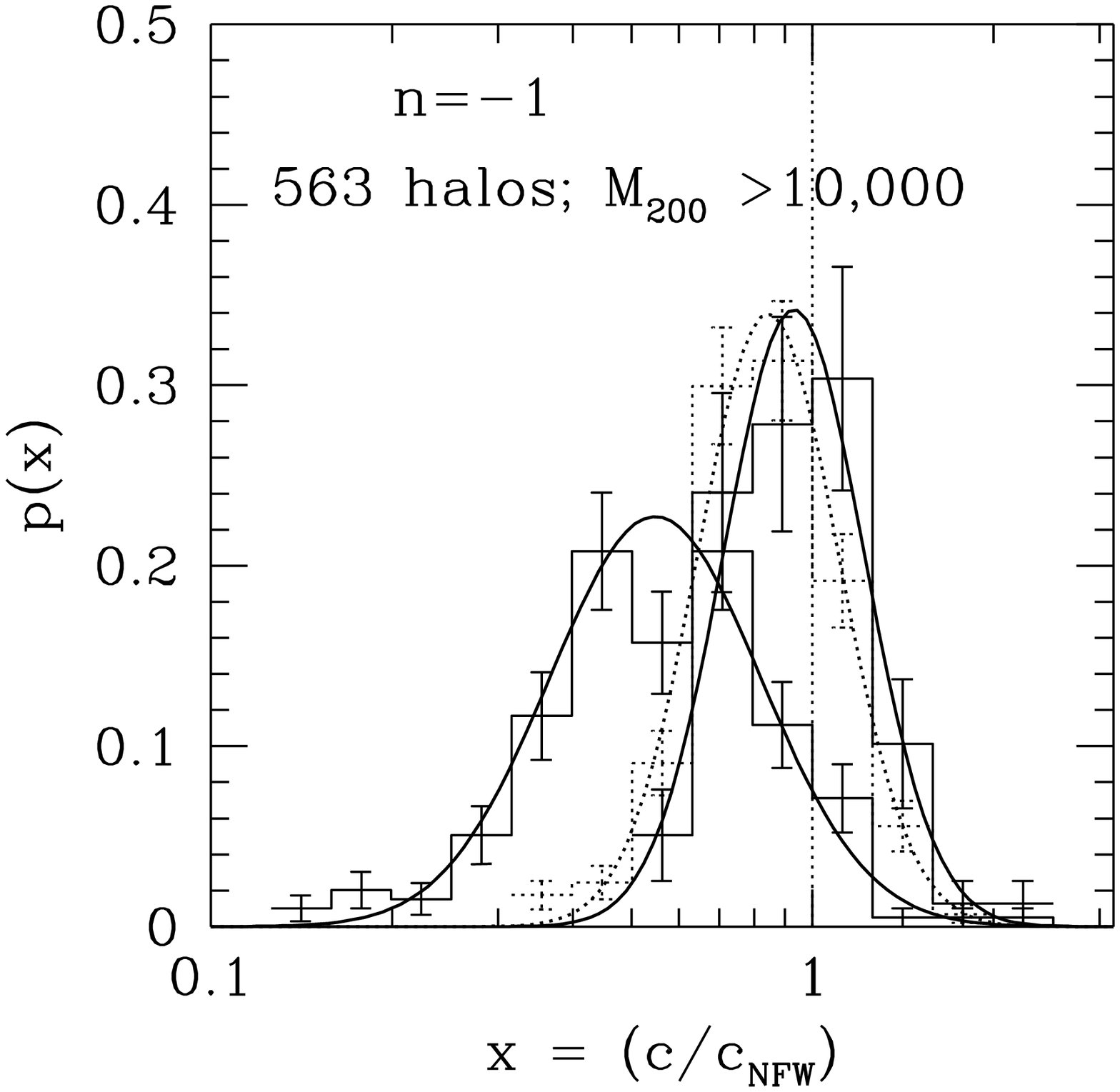,height=6cm}}
\caption[]{
  The probability distribution function of the concentration parameter
  $c$.  The right solid histogram is for the halos with
  $dvi_{max}<0.15$, the middle dotted one for $0.15<dvi_{max}<0.35$,
  and the left solid one for $dvi_{max}>0.35$. For each subset of the
  halos, the distribution of $c$ can be nicely fitted by a lognormal
  distribution.  }
\label{fig:1}
\end{figure*}

Recently Navarro, Frenk, \& White (1996,1998; NFW) have found that the
density profile $\rho(r)$ of the DM halos follows a universal form:
\begin{equation}
\rho(r)\propto {1\over r(r+1/c)^2}
\end{equation}
where $r$ is in units of the radius $R_{200}$ within which the
overdensity around the halo center is 200, and the density profile
depends only on the concentration parameter $c$. The parameter $c$ can
be predicted for each cosmogonic model with a recipe given by NFW, and
the predicted $c_{NFW}$ is a function of the halo mass only for a given
cosmogonic model (i.e. the power spectrum and the background
cosmology). These results are indeed very important and have many
interesting applications. However, it is important to point out that
NFW studied the density profile only for the halos {\it which look
in equilibrium}, so it is unclear how much fraction of the DM
halos really follows the universal form, since substructure is a common
phenomenon for DM halos in all viable cosmological models (e.g. Jing
et al. 1995). Even for halos which can be reasonably fit by the NFW
profile, a dispersion in $c$ seems inevitable because of 
a different formation history. Quantifying this
dispersion or more completely the Probability Distribution Function
(PDF) of $c$ is of considerable importance when applying the NFW
profile to interpret various observational facts. To answer these
important questions, many halos with good resolution (certainly much
more halos than that NFW had) are needed. Our simulations suit very well for
this purpose since typically one model (either CDM or SF) has 300 to
500 massive halos with at least 10,000 particles within the virial
radius.  Both the force and the mass resolutions of our simulations
are comparable to the NFW halos.

We have selected DM halos using the Friends-of-Friends method with the
linking length $0.2$ times of the mean particle separation. Then we
choose the local potential minimum as the center of the halo, 
and compute the density $\rho(r)$ in shells with
logarithmic thickness $\log_{10} \Delta r =0.1$ from $R_{200}$ inward
to $\eta$, the force resolution limit. The density profiles are then
fitted with Eq.~(1) to get the parameter $c$.  Although we have computed
the density profiles for the massive halos in all simulations, for the
convenience of the discussion in this very limited space (and talk
time), I present here the results only for one specific model, the
scale-free model of $n=-1$. A detailed discussion of all the models
will be given in a forthcoming paper (Jing \& Suto 1998, in preparation).

Firstly we note that the goodness of the fit varies strongly from halo
to halo.  Since the Poisson noise in the simulation density profile
is negligible, we define the goodness as the maximum relative
deviation of the simulation $\rho(r)$ from the fit $\rho_{NFW}(r)$ in
all the radial bins, i.e. $dvi_{max}=\max
\{|(\rho(r_i)-\rho^{NFW}(r_i))/ \rho^{NFW}(r_i)|\}$. About 35, 50
and 15 percent of the halos have the maximum deviations $dvi_{max}>35\%$,
$15\% <dvi_{max}<35\%$, and $dvi_{max}<15\%$ respectively.  With this
classification, it is fair to say that about 35\% halos could not be
fitted by the NFW profile, because of too significant substructures;
another 50\% halos with less substructures can be reasonably described
by the NFW profile, and the rest about 15\% halos with the least
substructures can be fitted by the NFW profile very nicely.

The PDF of the fitted $c$ is
presented in Figure 1. We choose $c/c_{NFW}$ as the abscissa in order
to correct for the mass dependence of the parameter $c$, though this
correction for the mass range covered by the halos here is actually
tiny. The PDF is shown separately for the halos with different amount
of substructures. For each subset of the halos, the PDF can be fitted
by a lognormal function
\begin{eqnarray}
p(c)dc={1\over \sqrt{2\pi}\sigma} &\nonumber\\
\times \exp{-{(\ln c-\ln \bar c)^2\over 2\sigma^2}} &d\ln c\,.
\end{eqnarray}
As we discussed in the last paragraph, because of the poor fit for the
halos with significant substructures, the PDF given by Fig.~1 for this
subset is probably not much meaningful and we shall not discuss
it. But for the other halos which can be reasonably fitted by the NFW
profile, the PDF is indeed a very important quantity. Both for the
most virliazed halos ($dvi_{max}<15\%$) and for the halos with
$15\%<dvi_{max}<35\%$, the dispersion $\sigma$ of the PDF is
0.27. The mean value $\bar c$ is $(0.93\pm 0.03)c_{NFW}$ for the halos
with $dvi_{max}<15\%$ and $(0.84\pm 0.02)c_{NFW}$ for the halos with
$15\%<dvi_{max}<35\%$. The PDF of the parameter $c$ depends weakly
on the cosmological parameters, and the above results can be applied
to other cosmological models with $\Omega_0\gs 0.25$. 

Our results are generally consistent with the results of NFW.
Variation of the density profiles was also found by  
Kravtsov et al. (1998) but was discussed in a
completely different way. What is new from our study is that we
have quantified how much halos can really be fitted by the NFW profile
and that we have derived the important quantity, the Probability
Distribution Function of the concentration parameter $c$, for the
halos which can be fitted by the NFW profile. These
results would be important for properly interpreting many cosmological
observations which are closely related to the density profile of
halos.

Before concluding this section, it would be important to point out
that because of the still limited resolution of out simulations, we
could not address the important problem that density profile might be
much steeper than $r^{-1}$ in the very inner region of the halos (
Fukushige \& Makino 1997; Moore et al. 1998). To definitely answer
this problem, one needs to simulate many halos with much higher
resolutions of both the mass and the force than the present work. Our work
in this direction is also in progress.

\section{The bias parameter of the dark matter halos}
\begin{figure*}
\centering\mbox{\psfig{figure=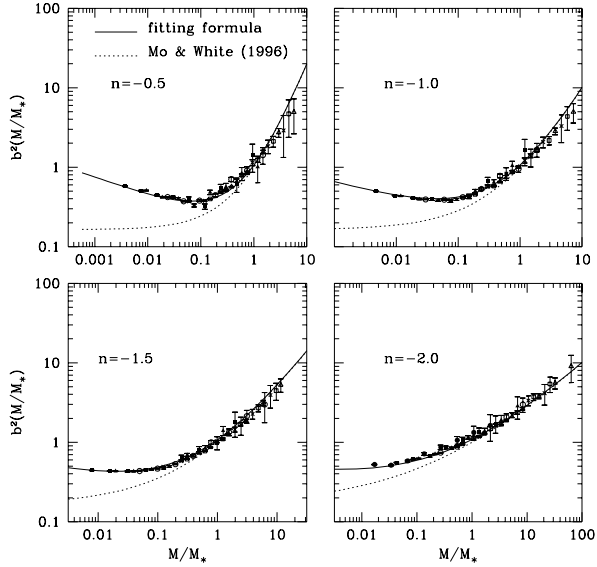,height=8cm}}
\caption[]{
The $b^2(M/M_{\star})$ for different evolution outputs
(different symbols). The dotted lines are the
analytical prediction of Mo \& White (1996), and the solid lines
are our fitting formula Eq.~3 which can
accurately fit the simulation results.  }
\label{fig:1}
\end{figure*}

The clustering of the DM halos is the basic block for modeling and
understanding the clustering of galaxies and of other extragalactic
objects. We have recently measured the two-point correlation function
of the DM halos for the scale free simulations with $n=-0.5$, $-1.0$,
$-1.5$, and $-2.0$ (Jing 1998). The results confirmed with
unprecedented accuracy that the bias of the halos $b$ is linear in the
linearly clustering regime. For a scale-free model, the bias parameter
$b$ is then expected to depend on the scaled mass $M/M_{\star}$ only,
where $M_{\star}$ is a characteristic non-linear mass. In Fig.~2, I
present the $b(M/M_{\star})$ for different simulation outputs of each
model.  The excellent scaling exhibited by the simulation data assures
that any numerical artifacts indeed have negligible effect on the
results of Fig.~2. The simulation results agree with the analytical
formula of Mo \& White (1996) only for massive halos with
$M/M_{\star}>1$, but are significantly higher for much less massive
halos. We also noted that there might be systematic difference between
the analytical prediction and our simulation results even at
$M/M_{\star}\gg 1$ (see Fig.~2), despite not with a high statistical
confidence. Our simulation results could be fitted very well by
\begin{eqnarray}
b(M)&=&({0.5\over \nu^4}+1)^{(0.06-0.02n)}\nonumber\\
&&\times (1+{\nu^2-1\over \delta_c})\,;\nonumber\\
\nu&=&(M/M_{\star})^{n+3/6}\,,
\end{eqnarray}
where $\delta_c=1.68$. This fitting formula
could be equally well applied to the CDM models with only a very simple
modification (Jing 1998). This finding has profound implications for the
clustering studies of galaxies. Further discussions are given in Jing (1998).

\section*{Acknowledgements} 
The author gratefully acknowledges Yasushi Suto for collaborations and
the JSPS foundation for a postdoctoral fellowship.  The simulations
were carried out on VPP/16R and VX/4R at NAOJ, Japan.

\end{document}